\documentclass[12pt,a4paper]{article}
\usepackage[dvips]{geometry}
\usepackage{amsmath}
\usepackage[dvips]{epsfig}

\DeclareMathOperator{\sech}{sech}
\DeclareMathOperator{\sn}{sn}
\DeclareMathOperator{\dn}{dn}
\DeclareMathOperator{\cn}{cn}

\begin{document}

\title{Analytical Multi-kinks in smooth potentials }
\author{G. P. de Brito\thanks{%
gustavopazzini@gmail.com}, R. A. C. Correa\thanks{
fis04132@gmail.com} and A. de Souza Dutra\thanks{
dutra@feg.unesp.br} \\
%EndAName
\textsl{UNESP-Campus de \ Guaratinguet\'{a}-DFQ, Av. Dr. Ariberto }\\
\textsl{Pereira Cunha, 333 C.P. 205 12516-410 Guaratinguet\'{a} SP Brasil}}
\maketitle

\begin{abstract}
In this work we present an approach which can be systematically used to
construct nonlinear systems possessing analytical multi-kink profile
configurations. In contrast with previous approaches to the problem, we are
able to do it by using field potentials which are considerably smoother than
the ones of Doubly Quadratic family of potentials. This is done without
losing the capacity of writing exact analytical solutions. The resulting
field configurations can be applied to the study of problems from condensed
matter to brane world scenarios.
\end{abstract}

\newpage

\section{Introduction}

\bigskip The study of nonlinear systems is growing since the sixties of the
last century \cite{b1,b2}. Nowadays the nonlinearity is found in many areas
of the physics, including condensed matter physics, field theory, cosmology
and others \cite{rajaramanweinberg}-\cite{b6.1}. Particularly, whenever we
have a potential with two or more degenerate minima, one can find different
vacua at different portions of the space. Thus, one can find domain walls
connecting such regions, and this necessarily leads to the appearance of
multi-kink configurations.

Some time ago, A. Champney and collaborators \cite{champney} made an
analysis of the reasons for the appearance of multi-kinks in dispersive
nonlinear systems. In part this was motivated by the discovery by Peyrard
and Kruskal \cite{a2} that a single kink becomes unstable when it moves in a
discrete lattice at sufficiently large velocity, whereas multi-kinks are
stable. The effect was shown to be associated with a resonant
interaction between the kink and the radiation \cite{a3}, and the resonances
were already observed experimentally \cite{a4}. In fact, in an earlier work by
Manton and Merabet \cite{manton}, it
was discussed the mechanism of production of kinks from excitations of the
internal mode, in a study of the dynamics of the interaction of two kinks
and one anti-kink in a $\phi^4$ model.  Furthermore, in a recent work by
M. A. Garcia-\~{N}ustes and J. A. Gonz\'{a}lez \cite{garcia}, it is shown
that a pair of kinklike solitons is emitted during the process of kink
breakup by internal mode instabilities in a Sine-Gordon model. The multi-kinks are
responsible, for instance, for a mobility hysteresis in a damped driven
commensurable chain of atoms \cite{a5}. Moreover in arrays of Josephson
junctions, instabilities of fast kinks generate bunched fluxon states
presenting multi-kink profiles \cite{a6}.

In a different context, by working with space-time dependent field
configurations, Coleman and collaborators \cite{coleman,coleman2} analyzed
the \textquotedblleft fate of the false vacuum\textquotedblright\ through a
semiclassical analysis of an asymmetric $\lambda \phi ^{4}$-like model.
There, they considered the decaying process of the field configuration from
the local to the global vacuum of the model . Moreover, some recent works
report fluctuating bouncing solutions \cite{dunne,plb rafael} in models
presenting local and global minima.

In all the above physical situations an analytical description of
multi-kinks would be very welcome. However, as far as we know, there is no
result in the literature which presents analytical multi-kink profiles
beyond the case of two, the so called double-kink \cite{pmodel}, \cite{physD}%
, \cite{hott}. Here we intend to fill this gap by presenting a general
procedure in order to construct analytical solutions for multi-kinks,
considering reasonably smooth field potentials.

In fact, the idea here is to improve another one used previously in some
works dealing with the so-called Double-Quadratic (DQ) model \cite{horovitz}%
, \cite{deleo}-\cite{losano}, whose potential is given by

\begin{equation}
V(\phi )=\frac{1}{2}\phi ^{2}-\left\vert \phi \right\vert +\frac{1}{2}.
\end{equation}

\noindent and some of their generalizations like the Asymmetrical
Double-Quadratic model (ADQ) \cite{b6} and the Generalized Asymmetrical
Double-Quadratic model (GADQ) \cite{plb rafael}. This last one has the
advantage that, having the previous mentioned potentials as their limits, it
can be used to study systems in which the curvature of the potential is
different in each side of the discontinuity point. Moreover, the vacua of
the model can be chosen to represent a kind of slow-roll potential, giving
rise to inflaton fields which are important in cosmological inflationary
scenarios. In fact, a similar model was used to study wet surfactant
mixtures of oil and water \cite{gompper}. The model is such that

\begin{equation}
V(\phi_{GADQ})=\left\{
\begin{array}{c}
\lambda\lbrack(\phi_{GADQ}-\phi_{2})^{2}+V_{2}],\text{ \ \ \ \ \ \ \ \ \ \ \
\ \ \ \ \ \ \ \ \ \ \ \ \ }\phi\geq0 \\
\\
\lambda\left( \frac{\phi_{2}^{2}+V_{2}}{\phi_{1}^{2}+V_{1}}\right)
[(\phi_{GADQ}+\phi_{1})^{2}+V_{1}],\text{ \ \ \ \ \ \ \ \ \ \ \ \ }\phi \leq0%
\text{\ }%
\end{array}
\right.  \label{adqg}
\end{equation}

\noindent where $\lambda$, $\phi_{1}$, $\phi_{2}$, $V_{1}$ and $V_{2}$ are
constant parameters which obey the following restrictions

\begin{equation}
\phi _{2}>0,\text{ \ }\phi _{1}>0,\text{ \ \ }V_{2}>-\phi _{2}^{2}\text{ \
and \ \ }V_{1}>-\phi _{1}^{2}.  \label{restrict}
\end{equation}

In this line of analysis, one can go further by studying a model with a
potential for the scalar field presenting four degenerate minima, for
instance. This model is written as%
\begin{equation}
V(\phi )=\left\{
\begin{array}{l}
\frac{\lambda _{1}}{2}\left[ \left( \phi +\frac{3a}{2}\right) ^{2}+b_{1}%
\right] ;\qquad -\infty <\phi \leq -a, \\
\frac{\lambda _{2}}{2}\left( \frac{a^{2}+4b_{1}}{a^{2}+4b_{2}}\right) \left[
\left( \phi +\frac{a}{2}\right) ^{2}+b_{2}\right] ;\qquad -a\leq \phi \leq 0,
\\
\frac{\lambda _{3}}{2}\left( \frac{a^{2}+4b_{1}}{a^{2}+4b_{3}}\right) \left[
\left( \phi -\frac{a}{2}\right) ^{2}+b_{3}\right] ;\qquad 0\leq \phi \leq a,
\\
\frac{\lambda _{4}}{2}\left( \frac{a^{2}+4b_{1}}{a^{2}+4b_{4}}\right) \left[
\left( \phi -\frac{3a}{2}\right) ^{2}+b_{4}\right] ;\qquad a\leq \phi
<\infty .%
\end{array}%
\right.
\end{equation}

\noindent where%
\begin{equation}
\lambda _{2}=\lambda _{1}\left( \frac{a^{2}+4b_{1}}{a^{2}+4b_{2}}\right) ,%
\text{ }\lambda _{3}=\lambda _{1}\left( \frac{a^{2}+4b_{1}}{a^{2}+4b_{3}}%
\right) ,\text{ }\lambda _{4}=\lambda _{1}\left( \frac{a^{2}+4b_{1}}{%
a^{2}+4b_{4}}\right) .
\end{equation}

Notice that, despite the fact that the potential is continuous by parts,
their physically acceptable solutions must be continuous both for the field
as for its first derivative. This restriction comes from the fact that one
should seek for configurations where the energy density is a continuous and
non-singular one. Thus one must impose that the solutions are such that $%
\phi \left( x\right) $ and $d\phi \left( x\right) /dx$ are continuous
throughout the spatial axis. In fact, in this class of potentials one only
needs to require that $\left(\frac{d\Phi}{dx}\right)^2 $ be continuous due
to the continuity of the energy density. In Figure (1) we can see the
profile of the above potential, as well as its triple-kink configuration.

By following the approach developed in \cite{plb rafael} one can determine a
solution which presents four regions where the fields are connecting the
different (local or global) vacua of the model. Specifically, we look for a
solution which is continuous and have its first spatial derivative also
continuous, where $\phi ^{(1)}(x)\rightarrow -3a/2$ at $x\rightarrow -\infty
$ and $\phi ^{(4)}(x)\rightarrow 3a/2$ at $x\rightarrow \infty $. After
straightforward calculations one can verify that the solution is given by%
\begin{equation}
\phi (x)=%
\begin{cases}
-\frac{3a}{2}+B_{1}e^{x\sqrt{\lambda _{1}}} & -\infty <x<-a,~ \\
-\frac{a}{2}+A_{2}e^{-x\sqrt{\lambda _{2}}}+B_{2}e^{x\sqrt{\lambda _{2}}} &
-a<x<0, \\
\frac{a}{2}+A_{3}e^{-x\sqrt{\lambda _{3}}}+B_{3}e^{x\sqrt{\lambda _{3}}} &
0<x<a, \\
\frac{3a}{2}+A_{4}e^{-x\sqrt{\lambda _{4}}} & \quad a<x<\infty ,%
\end{cases}%
\end{equation}

\noindent where $A^{\prime }s$ and $B^{\prime }s$ are integrating constants.
In order to ensure the necessary gluing conditions at each junction point,
one arrive to the following expressions for the parameters in $\phi \left(
x\right) $

\begin{eqnarray}
B_{1} &=&\frac{a\sqrt{\text{$\lambda $}_{2}}e^{a\sqrt{\text{$\lambda $}_{1}}%
}[2\cosh (\sqrt{\text{$\lambda $}_{2}}a)+1]}{2[\sqrt{\text{$\lambda $}_{2}}%
\cosh \left( \sqrt{\text{$\lambda $}_{2}}a\right) +\sqrt{\text{$\lambda $}%
_{1}}\sinh \left( \sqrt{\text{$\lambda $}_{2}}a\right) ]},~  \notag \\
A_{2} &=&\frac{a[\sqrt{\text{$\lambda $}_{2}}+\sqrt{\lambda _{1}}(1+2e^{-a%
\sqrt{\text{$\lambda $}_{2}}}]}{2[\sqrt{\text{$\lambda $}_{2}}+\sqrt{\text{$%
\lambda $}_{1}}+(\sqrt{\text{$\lambda $}_{2}}-\sqrt{\text{$\lambda $}_{1}}%
)e^{-2a\sqrt{\text{$\lambda $}_{2}}}]},  \notag \\
B_{2} &=&-\frac{2a\sqrt{\text{$\lambda $}_{1}}+a(\sqrt{\text{$\lambda $}_{1}}%
-\sqrt{\text{$\lambda $}_{2}})e^{-a\sqrt{\text{$\lambda $}_{2}}}}{4[\sqrt{%
\text{$\lambda $}_{2}}\cosh \left( \sqrt{\text{$\lambda $}_{2}}a\right) +%
\sqrt{\text{$\lambda $}_{1}}\sinh \left( \sqrt{\text{$\lambda $}_{2}}%
a\right) ]}, \\
~A_{3} &=&\frac{a[-\sqrt{\text{$\lambda $}_{3}}+\sqrt{\lambda _{4}}(1+2e^{a%
\sqrt{\text{$\lambda $}_{3}}}]}{2[\sqrt{\text{$\lambda $}_{3}}-\sqrt{\text{$%
\lambda $}_{4}}+(\sqrt{\text{$\lambda _{4}$}}+\sqrt{\text{$\lambda $}_{3}}%
)e^{2a\sqrt{\text{$\lambda $}3}}]},  \notag \\
B_{3} &=&\frac{a}{2}\left[ \frac{\sqrt{\text{$\lambda $}_{3}}-\sqrt{\lambda
_{4}}(1+2e^{a\sqrt{\text{$\lambda $}_{3}}}]}{\sqrt{\text{$\lambda $}_{3}}-%
\sqrt{\text{$\lambda $}_{4}}+(\sqrt{\text{$\lambda $}_{3}}+\sqrt{\text{$%
\lambda $}_{4}})e^{2a\sqrt{\text{$\lambda $}_{3}}}}\right] ,~  \notag \\
A_{4} &=&\frac{-a\sqrt{\text{$\lambda $}_{3}}e^{a\sqrt{\text{$\lambda $}_{4}}%
}[2\cosh (\sqrt{\text{$\lambda $}_{3}}a)+1]}{2[\sqrt{\text{$\lambda $}_{3}}%
\cosh \left( \sqrt{\text{$\lambda $}_{3}}a\right) +\sqrt{\text{$\lambda $}%
_{4}}\sinh \left( \sqrt{\text{$\lambda $}_{3}}a\right) ]}.  \notag
\end{eqnarray}

\bigskip However, in all these models one has potentials which are
discontinuous in the first derivative with respect to the field. This is the
price to be paid in order to assure that the second order derivatives
describing the evolution of the field configuration be linear by parts, and
the non-linear effect comes precisely from that discontinuity.

Notwithstanding, we do not know how to solve analytically only linear
equations but some non-linear too. For instance, we can get exact analytical
solutions for the so called $\lambda \phi ^{4}$ and the sine-Gordon models.

The principal idea to be developed in this work is to use this ability,
associated with the possibility of constructing a smooth multi-degenerate
minima potential by gluing the potentials of many $\lambda \phi ^{4}$
models, or Sine-Gordon and $\lambda \phi ^{4}$ models, for different ranges
of the scalar field. Once the potential is constructed, one can use an
absolutely analogous procedure to that used for the case of the DQ model and
his fellows. As a consequence, we develop in this work an approach which
allow us to construct models with multi-step kinks, due to the presence of a
chosen number of intermediate local vacua, in considerably smooth potentials.

\section{Models involving $n$ local vacua}

\bigskip \indent As previously asserted, we present here a general approach
capable of describing analytical configurations with an arbitrary number of
\ local vacua and, as a consequence, presenting a multi-kink profile. In
this work we present some models which, as far as we know, are new in the
literature and are examples of a wider class of models which can be
analytically solved in order to construct field configurations with an
arbitrary number of \textquotedblleft steps" in a general multi-kink
profile. The potentials which describe these models present two global vacua
($V(\phi _{M})=0$) and $n$ local ones ($V(\phi _{m})>0$). As asserted in the
Introduction section, they allow one to obtain multi-kink analytical
solutions in smooth potentials.

\indent The first model considered here is the one defined as%
\begin{equation}
V(\phi )=%
\begin{cases}
\frac{\lambda ^{2}}{2}[a^{2}-(\phi -nb)^{2}]^{2} & \text{; $\phi \geq nb$}
\\
&  \\
\frac{\alpha ^{2}}{2}B^{2}-\frac{\alpha ^{2}}{2}[b^{2}-(\phi -kb)^{2}]^{2} &
\text{; $(k-1)b\leq \phi <(k+1)b$} \\
&  \\
\frac{\lambda ^{2}}{2}[a^{2}-(\phi +nb)^{2}]^{2} & \text{; $\phi <-nb$},%
\end{cases}
\label{10}
\end{equation}%
where, for even $n$, $k$ shall assume all odd values inside the interval $%
[1-n,n-1]$, and for odd values of $n$, $k$ will assume all the even values
along the same interval. In other words, $n$ defines the number of local
vacua and $k$ labels each one of them. We will denote the global vacua as $%
\phi _{M}$, and by $\phi _{k}$ the local ones. Furthermore, the matching
points will be denoted by $\phi _{J}$. For the case presented in this
section, the global vacua will be localized at the points where $\phi
_{M}=\pm (a+nb)$, while the local ones will be found at $\phi _{k}=kb$ . In
order to get smooth connections, the stepwise potential that we study here
was constructed through the junction of polynomial potentials of degree
four, the so called $\phi ^{4}$ model, which in each region was conveniently
dislocated. In the border regions the shift was given by $\phi \rightarrow
\phi \pm nb$, while in the intermediate regions ($(k-1)b\leq \phi <(k+1)$)
it was used the inverted $\phi ^{4}$ model. The constant which is added at
each intermediate region, is conveniently chosen in order to keep the global
vacua at the extremals equal to zero and, as a consequence, we must to impose
that $B>b^{2}$ in order grant that these vacua are really local. Since we
are dealing with a stepwise potential, some matching conditions must be
imposed at the border $\phi _{J}$ of each region. For instance we shall have
\begin{equation}
\lim_{\phi \rightarrow \phi _{J}^{-}}V(\phi )=\lim_{\phi \rightarrow \phi
_{J}^{+}}V(\phi ),  \label{junction}
\end{equation}%
\noindent which implies into the constraint
\begin{equation}
\lambda a^{2}=\alpha B.  \label{constantes}
\end{equation}%
It is interesting to note that this relation between $\alpha $ and $\lambda $
is also enough to keep the derivative of the potential continuous at the
junctions.

\indent The second model which we will consider in this work is such that
one has%
\begin{equation}
V(\phi )=%
\begin{cases}
\frac{\lambda ^{2}}{2}[a^{2}-(\phi -\frac{n\pi }{\mu })^{2}]^{2} & \text{; $%
\phi \geq \frac{n\pi }{\mu }$} \\
&  \\
\frac{\alpha ^{2}}{2}\left[ B^{2}+\frac{1+(-1)^{n}\cos \left( \mu \phi
\right) }{2}\right] & \text{; $|\phi |<\frac{n\pi }{\mu }$} \\
&  \\
\frac{\lambda ^{2}}{2}[a^{2}-(\phi +\frac{n\pi }{\mu })^{2}]^{2} & \text{; $%
\phi \leq \frac{-n\pi }{\mu }$},%
\end{cases}
\label{potecial_2}
\end{equation}%
once more, $n$ is an integer number corresponding to the number of local
vacua of the potential. In this case we construct the model by gluing two $%
\phi ^{4}$ potentials, located at the border regions, with a sine-Gordon
potential which will be responsible for the intermediary local vacua. In
this model, the global vacua are at $\phi _{M}=\pm (a+n\pi /\mu )$, while
the local ones are at the points where $\phi _{k}=k\pi /\mu $. For even $n$,
$k$ assume all the odd numbers along the interval $[1-n,n-1]$, and for odd $%
n $, $k$ assume all the even values at the same interval. Now, the
continuity condition leads to the following constraint between the coupling
constants of the potential
\begin{equation}
\lambda a^{2}=\alpha \sqrt{1+B^{2}}.  \label{constantes_2}
\end{equation}%
As in the previous model, this condition is also enough to keep the
derivative of the potential continuous. As we are going to see below, the
relation (\ref{constantes_2}) is capable to grant the continuity of the
field configuration, assuring that the energy of the configuration stays
finite.

\section{Solution for the first smooth potential with symmetric local vacua:}

\bigskip

\indent In this section we develop the solutions and analyze some features
of the first model here proposed, the one defined in (\ref{10}). We are
primordially interested in solutions connecting the global vacua of the
potential. These solutions obey the following boundary conditions
\begin{equation}
\lim_{x\rightarrow \pm \infty }\phi (x)=\pm (a+nb).  \label{boundary_1}
\end{equation}

Since we are working with the usual Lagrangian density for a
self-interacting scalar field, the corresponding equation of motion is
\begin{equation}
\partial _{\mu }\partial ^{\mu }\phi +\frac{dV(\phi )}{d\phi }=0.
\label{eq_mov}
\end{equation}%
But, we are looking for the static solutions, which can be boosted in order
to recover the traveling ones. So, the above equation is simplified to
\begin{equation}
\phi ^{\prime \prime }=\frac{dV(\phi )}{d\phi },
\end{equation}%
which can be integrated to
\begin{equation}
\frac{1}{2}\phi ^{\prime 2}=V(\phi ).  \label{first_order}
\end{equation}%
It must be noticed that the corresponding integration constant was chosen as
zero, since we are interested in configurations which go asymptotically to
some vacuum of the field potential.

\indent From the equation (\ref{first_order}), as usual, we can write
\begin{equation}
\frac{d\phi }{dx}=\pm \sqrt{2V(\phi )},
\end{equation}%
and from the above one we can conclude that the solution for $\phi (x)$ will
be monotonically growing or decreasing according the chosen sign of the
right hand side of the equation. On the other hand, we are interested in
solutions going from the negative vacuum to the positive one, so we will
look for a field which increases with $x$.

\indent As we have seen before, $V(\phi )$ is a continuous by parts
function, specifically the potential has $n+2$ distinct regions, each one
governed by one differential equation coming from (\ref{first_order}). So,
in order to identify the solutions and their respective regions we will
label them as $\phi _{i}(x)$, with $1\leq i\leq n+2$ coming from the left to
the right.

\indent There is a question which must be carefully treated, and that is the
case of the domain of the solutions $\phi (x)$, since each region in the
space of the fields corresponds to another in the coordinate space. In order
to become the analysis more precise we relate each region $i$ of the
potential $V(\phi )$ to a set $\Phi _{i}$ such that $\phi _{i}(x)\in \Phi
_{i}$. Correspondingly to each set $\Phi _{i}$ we have a set $X_{i}$, so
that we can associate each element $x\in X_{i}$ to an element of $\ \Phi
_{i} $ through the map $\phi :X_{i}\longrightarrow \Phi _{i}$. In fact, each
set $\Phi _{i}$ is well determined due to the definition of the potential $%
V(\phi )$. However, it is still necessary to determine the sets $X_{i}$. For
the first region of the potential, $\Phi _{1}=\{\phi \quad |-\infty <\phi
<-nb\}$, the boundary condition (\ref{boundary_1}) leads to $x\rightarrow
-\infty \Rightarrow \phi (x)\rightarrow -(a+nb)$. Furthermore, since $\phi
(x)$ is a growing function, there is a certain value of $x_{1}$ such that $%
\phi (x_{1})=-nb$ , in such a way that if $x<x_{1}$ then $\phi (x)\in \Phi
_{1}$ and for $x\geq x_{1}$ the solution $\phi (x)$ must be in the second
region of the potential. Thus, we can conclude that $X_{1}=\{x\quad |-\infty
<x<x_{1}\}$. In the second region of the potential, $\Phi _{2}=\{\phi \quad
|-nb\leq \phi <(2-n)b\}$, we can use the same argument in order to conclude
that $X_{2}=\{x\quad |x_{1}\leq x<x_{2}\}$, with $\phi (x_{2})=(2-n)b$ and
for $x\geq x_{2}$ the solution $\phi $ corresponds to third region of the
potential and so on. Generally speaking, we can write for the intermediary
regions of the potential $\Phi _{i}=\{\phi \quad |(k-1)b\leq \phi <(k+1)b\}$%
, and the domain of $\phi (x)$ can be defined as $X_{i}=\{x\quad
|x_{i-1}\leq x<x_{i}\}$. Finally, for the right-hand border of the potential
we have $\Phi _{n+2}=\{\phi \quad |+nb\leq \phi <+\infty \}$ and the domain
in the coordinate space will be given by $X_{n+2}=\{x\quad |x_{n+1}\leq
x<+\infty \}$.

\indent\textbf{Solution in the first region:} Here we are interested in
solutions like $\phi :X_{1}\longrightarrow \Phi _{1}$. In this region the
corresponding differential equation is given by
\begin{equation}
\frac{1}{2}\phi ^{\prime 2}=\frac{\lambda ^{2}}{2}(a^{2}-(\phi +nb)^{2})^{2},
\label{reg_1}
\end{equation}%
performing the translation $\varphi =\phi +nb$, the above equation can be
cast in the form
\begin{equation}
\frac{1}{2}\varphi ^{\prime 2}=\frac{\lambda ^{2}}{2}(a^{2}-\varphi
^{2})^{2},
\end{equation}%
and this last equation corresponds to the usual one for the $\varphi ^{4}$
model, and can be easily integrated to give $\varphi (x)=a\tanh (\lambda
a(x-x_{0}))$. Returning to the original variable, one obtains
\begin{equation}
\phi (x)=a\tanh (\lambda a(x-x_{1}))-nb\quad ;~-\infty <x<x_{1},
\label{solution_1}
\end{equation}%
which satisfies the adopted boundary conditions.

\indent\textbf{Solution in the region n+2:} Focusing our attention in the
right border of the potential, where we seek for a solution like $\phi
:X_{n+2}\longrightarrow \Phi _{n+2}$, we can see that
\begin{equation}
\frac{1}{2}\phi ^{\prime 2}=\frac{\lambda ^{2}}{2}(a^{2}-(\phi -nb)^{2})^{2},
\label{reg_2}
\end{equation}%
again we can perform a displacement n as did in above $\varphi =\phi -nb$
and this leads to
\begin{equation}
\phi (x)=a\tanh (\lambda a(x-x_{n}))+nb\quad ;x_{n+1}\leq x<+\infty .
\label{solution_2}
\end{equation}%
Once more the boundary conditions are respected as required.

\indent\textbf{Solutions in the intermediary regions:} Let us now deal with
the case where $\phi :X_{i}\longrightarrow \Phi _{i}$. In those regions, the
differential equations are given by
\begin{equation}
\frac{1}{2}\phi ^{\prime 2}=\frac{\alpha ^{2}}{2}B^{2}-\frac{\alpha ^{2}}{2}%
[b^{2}-(\phi -kb)^{2}]^{2},
\end{equation}%
and, in this case, we need a bit more manipulation before getting the
solution. First of all, we rewrite the equation in the form
\begin{equation}
\frac{d\varphi }{dx}=\alpha \sqrt{(A-\varphi ^{2})(\varphi ^{2}-\tilde{A})},
\end{equation}%
where $A=b^{2}+B$, $\tilde{A}=b^{2}-B$ and $\varphi =\phi -kb$. Now, making
the transformation $\varphi =-\sqrt{A}\cos \theta $, it can be rewritten as
\begin{equation}
\frac{d\theta }{dx}=\sqrt{\frac{2}{B}}\lambda a^{2}\sqrt{1-m\sin ^{2}\theta }%
,
\end{equation}%
where we used the equation (\ref{constantes}) and defined $m=\frac{A}{A-%
\tilde{A}}=\frac{b^{2}+B}{2B}$. Then, integrating it between $x_{i-1}$ and $%
x $ we get
\begin{equation}
\int_{\theta (x_{i-1})}^{\theta (x)}\frac{d\theta ^{\prime }}{\sqrt{1-m\sin
^{2}\theta ^{\prime }}}=\sqrt{\frac{2}{B}}\lambda
a^{2}\int_{x_{i-1}}^{x}dx^{\prime }=\sqrt{\frac{2}{B}}\lambda
a^{2}(x-x_{i-1}).  \label{int_1}
\end{equation}%
The left-hand side of (\ref{int_1}) can be identified with an elliptical
integral and, provided that $m\in \lbrack 0,1]$, it can be solved in terms
of Jacobi elliptic functions \cite{b8}. More precisely we have
\begin{equation}
\int_{\theta (x_{i-1})}^{\theta (x)}\frac{d\theta ^{\prime }}{\sqrt{1-m\sin
^{2}\theta ^{\prime }}}=\sn^{-1}(\sin \theta |m)\bigg|_{\theta
(x_{i-1})}^{\theta (x)}.  \label{elliptic}
\end{equation}%
Substituting this result in (\ref{int_1}), and after some manipulation, we
arrive at
\begin{equation}
\sin \theta (x)=\sn\bigg(\sn^{-1}[\sin \theta (x_{i-1})]+\sqrt{\frac{2}{B}}%
\lambda a^{2}(x-x_{i-1})\bigg|\frac{b^{2}+B}{2B}\bigg).
\end{equation}%
Finally, returning to the original variables we get
\begin{equation}
\phi (x)=\sqrt{b^{2}+B}\cn\bigg(\delta _{i-1}+\sqrt{\frac{2}{B}}\lambda
a^{2}(x-x_{i-1})\bigg|\frac{b^{2}+B}{2B}\bigg)+kb\quad ;x_{i-1}\leq x<x_{i},
\end{equation}%
where $\sn(u|m)$ represents the elliptic sine function, $\cn(u|m)$
represents the elliptic cosine and we defined that $\delta _{i-1}=\sn%
^{-1}[\sin \theta (x_{i-1})]$.

\indent Now, using the relation $\phi -kb=-\sqrt{A}\cos \theta $ and also
the continuity conditions at the junction points, which are given by $\phi
(x_{i-1})=(k-1)b$, it can be \ verified that $\delta _{i-1}$ must satisfy
the following constraint equation
\begin{equation}
\cn\bigg(\delta _{i-1}\bigg|\frac{b^{2}+B}{2B}\bigg)+\frac{b}{\sqrt{b^{2}+B}}%
=0.  \label{delta}
\end{equation}%
It can be verified from above that $\delta _{i-1}$ is independent from the
chosen region of the potential, which allow us to use the notation $\delta
\equiv \delta _{i-1}$. However, from the equation (\ref{delta}) we can
conclude that $\delta $ can not be univocally determined, since there is an
infinite number of solutions for $\delta $ due to the periodicity of the
elliptic cosine. On the other hand, despite the fact that the solutions of (%
\ref{delta}) are enough to keep the continuity of $\phi (x)$, not all of
them grant the continuity of its derivative at the junction points.
Furthermore, it can be numerically observed that half of the roots of (\ref%
{delta}) lead to continuous $\phi (x)$, while the other half lead to
continuous $\phi ^{\prime }(x)$, and this behavior under the continuity of
the solutions is alternate. Fortunately, it can be numerically verified that
the first negative root of the equation (\ref{delta}) is able to grant the
continuity of $\phi (x)$ and its derivative. A further important point is
the one related to the determination of the position of the junction points $%
x_{i}$. In fact, $x_{1}$ can be arbitrarily chosen due to the translation
symmetry inherent to the model. However, the points $x_{2}$, $x_{3}$, ..., $%
x_{i}$ depend of the choice of $x_{1}$. Once more using the boundary
conditions at the junction points, we get the constraint relation
\begin{equation}
\cn\bigg(\delta +\sqrt{\frac{2}{B}}\lambda a^{2}(x_{i}-x_{i-1})\bigg|\frac{%
b^{2}+B}{2B}\bigg)=\frac{b}{\sqrt{b^{2}+B}}.  \label{xi1}
\end{equation}%
Once $x_{1}$ is specified, we can use the above equation in order to
determine $x_{2}$. It can be seen from the above equation that $%
x_{i}-x_{i-1} $ is independent of the region of the potential that is under
study, so we can verify that
\begin{equation}
x_{2}-x_{1}=x_{3}-x_{2}=x_{4}-x_{3}=...=x_{i}-x_{i-1}=...=x_{n+1}-x_{n}.
\label{xi2}
\end{equation}%
By using this last equation, we can still conclude that one can write
\begin{equation}
x_{i}=(i-1)x_{2}-(i-2)x_{1},
\end{equation}%
so, given a $x_{1}$, the remaining $x_{i}$ can be computed from the
equations (\ref{xi1}) and (33).

\indent In general we can write the solution of this model in the form
\begin{equation}
\phi (x)=%
\begin{cases}
a\tanh (\lambda a(x-x_{1}))-nb & \quad ;-\infty <x<x_{1} \\
&  \\
\sqrt{b^{2}+B}\cn\bigg(\delta +\sqrt{\frac{2}{B}}\lambda a^{2}(x-x_{i-1})%
\bigg|\frac{b^{2}+B}{2B}\bigg)+kb & \quad ;x_{i-1}\leq x<x_{i} \\
&  \\
a\tanh (\lambda a(x-x_{n+1}))+nb & \quad ;x_{n+1}\leq x<+\infty .%
\end{cases}
\label{phi_1}
\end{equation}%
As it was mentioned above, the parameter $B$ controls the height of the
local vacua in the model and, as one can verify, when $B$ approaches its
critical value $b^{2}$, the intermediary solutions assume a kink-like
profile, and the complete solution behaves like a kind of multi-kink, like a
ladder with many steps and this, as far as we know, is new in terms of an
analytical configuration. In the limit case where $B=b^{2}$, we find a
configuration where the potential exhibits only global vacua and, as a
consequence, the solution connects the adjacent vacua through kink-like
configurations. \ The field configurations which obey the equation (\ref%
{first_order}) have their energy density given by
\begin{equation}
\varepsilon \lbrack \phi (x)]=\bigg(\frac{d\phi (x)}{dx}\bigg)^{2},
\label{energy}
\end{equation}%
and this leads to the following energy density for the present case
\begin{equation}
\varepsilon \lbrack \phi (x)]=%
\begin{cases}
\lambda ^{2}a^{4}\sech^{2}(\lambda a(x-x_{1})) & \,\,;-\infty <x<x_{1} \\
&  \\
\frac{2\lambda ^{2}a^{4}}{B}(b^{2}+B)\cn^{2}\bigg(h(x)\bigg|\frac{b^{2}+B}{2B%
}\bigg)\dn^{2}\bigg(h(x)\bigg|\frac{b^{2}+B}{2B}\bigg) & \quad ;x_{i-1}\leq
x<x_{i} \\
&  \\
\lambda ^{2}a^{4}\sech^{2}(\lambda a(x-x_{n+1})) & \quad ;x_{n+1}\leq
x<+\infty ,%
\end{cases}
\label{energy_phi_1}
\end{equation}%
where $h(x)=\delta +\sqrt{2/B}\lambda a^{2}(x-x_{i-1})$ and $\dn(u,m)$ is a
Jacobi function. In fact, the energy density above obtained is an integrable
function which keeps the total energy finite as necessary. In Figure 2, it
is presented a typical example of this first situation, including the
potential, the corresponding triple kink and its energy density.

\section{Asymmetric version of a model with local vacua:}

\indent In this section we work with an asymmetrical potential presenting
two local vacua and which can be described by the stepwise function
\begin{equation}
V(\phi )=%
\begin{cases}
\frac{\lambda _{1}^{2}}{2}[a_{1}^{2}-(\phi -2b_{1})^{2}]^{2} & \text{; $\phi
\geq 2b_{1}$} \\
&  \\
\frac{\alpha _{1}^{2}}{2}B_{1}^{2}-\frac{\alpha _{1}^{2}}{2}[b_{1}^{2}-(\phi
-b_{1})^{2}]^{2} & \text{; $0\leq \phi <2b_{1}$} \\
&  \\
\frac{\alpha _{2}^{2}}{2}B_{2}^{2}-\frac{\alpha _{2}^{2}}{2}[b_{2}^{2}-(\phi
+b_{2})^{2}]^{2} & \text{; $-2b_{2}\leq \phi <0$} \\
&  \\
\frac{\lambda _{2}^{2}}{2}[a_{2}^{2}-(\phi +2b_{2})^{2}]^{2} & \text{; $\phi
<-2b_{2}$}.%
\end{cases}
\label{potecial_3}
\end{equation}%
This model was constructed from the potential ($\ref{10}$), However, for the
sake of simplicity, we considered a version with only two local vacua, but
it is important to remark that it can be extended in order to have an
arbitrary number of local vacua. Thus, like in the symmetric case, we must
to keep it continuous at the junction points and, using the equation (\ref%
{junction}), we get the following constraint between their coupling
constants
\begin{equation}
\lambda _{1}a_{1}^{2}=\lambda _{2}a_{2}^{2}=\alpha _{1}B_{1}=\alpha
_{2}B_{2}.
\end{equation}%
One can verify that the conditions $B_{1}>b_{1}^{2}$ and $B_{2}>b_{2}^{2}$
are also necessary to grant the presence of local vacua. The solution of the
differential equation (\ref{first_order}) in each region of the potential (%
\ref{potecial_3}) follows the same line of reasoning as in the previous case
and, due to this, we will present the solution directly in this section. So,
in this case we have
\begin{equation}
\phi (x)=%
\begin{cases}
a_{1}\tanh (\lambda _{1}a_{1}(x-x_{1}))-2b_{1} & \quad ;-\infty <x<x_{1} \\
&  \\
\sqrt{b_{1}^{2}+B_{1}}\cn\bigg(\delta _{1}+\sqrt{\frac{2}{B_{1}}}\lambda
_{1}a_{1}^{2}(x-x_{1})\bigg|\frac{b_{1}^{2}+B_{1}}{2B_{1}}\bigg)-b_{1} &
\quad ;x_{1}\leq x<x_{2} \\
&  \\
\sqrt{b_{2}^{2}+B_{2}}\cn\bigg(\delta _{2}+\sqrt{\frac{2}{B_{2}}}\lambda
_{1}a_{1}^{2}(x-x_{2})\bigg|\frac{b_{2}^{2}+B_{2}}{2B_{2}}\bigg)+b_{2} &
\quad ;x_{2}\leq x<x_{3} \\
&  \\
a_{2}\tanh \bigg(\lambda _{1}\frac{a_{1}^{2}}{a_{2}}(x-x_{3})\bigg)+2b_{2} &
\quad ;x_{3}\leq x<+\infty .%
\end{cases}
\label{phi_assimetrico}
\end{equation}%
The continuity of $\phi (x)$, makes necessary that the constants $\delta
_{1} $ and $\delta _{2}$ satisfy the following conditions
\begin{align}
\cn\bigg(\delta _{1}\bigg|\frac{b_{1}^{2}+B_{1}}{2B_{1}}\bigg)+\frac{b_{1}}{%
\sqrt{b_{1}^{2}+B_{1}}}& =0,  \notag  \label{delta1} \\
\cn\bigg(\delta _{2}\bigg|\frac{b_{2}^{2}+B_{2}}{2B_{2}}\bigg)+\frac{b_{2}}{%
\sqrt{b_{2}^{2}+B_{2}}}& =0.
\end{align}%
As it can be seen, in contrast with the case of the equation (\ref{delta}),
in this case the $\delta $ parameters depend on the region of the potential.
Furthermore, as in the previous case, there are many solutions for $\delta $
satisfying the above equations. Once more it can be checked numerically that
the first negative solutions warrant the continuity of $\phi (x)$ and its
derivative. We can also see how $x_{1}$, $x_{2}$ and $x_{3}$ are related to
each other. As in the previous case $x_{1}$ can be arbitrarily chosen and,
once it is defined, the remaining coordinates, $x_{2}$ and $x_{3} $, can be
obtained through the equations
\begin{align}
\cn\bigg(\delta _{1}+\sqrt{\frac{2}{B_{1}}}\lambda _{1}a_{1}^{2}(x_{2}-x_{1})%
\bigg|\frac{b_{1}^{2}+B_{1}}{2B_{1}}\bigg)& =\frac{b_{1}}{\sqrt{%
b_{1}^{2}+B_{1}}},  \notag  \label{x2} \\
\cn\bigg(\delta _{2}+\sqrt{\frac{2}{B_{2}}}\lambda _{1}a_{1}^{2}(x_{3}-x_{2})%
\bigg|\frac{b_{2}^{2}+B_{2}}{2B_{2}}\bigg)& =\frac{b_{2}}{\sqrt{%
b_{2}^{2}+B_{2}}}.
\end{align}%
In fact, each one of the above equations is very similar to the one in (\ref%
{xi1}), but in the asymmetrical case there is no expression which is
analogous to (\ref{xi2}).

In this model, the parameters which are responsible for controlling the
local vacua are $B_{1}$ and $B_{2}$, and when they approach respectively
their critical values $b_{1}^{2}$ and $b_{2}^{2}$, the intermediary
solutions start to present the expected typical kink profile. In the
particular case where $n=2$, the solution obtained is an asymmetrical triple
kink, as one can see in the Figure 3. Finally, through the equation (\ref%
{energy}) it can be computed the energy density, which is given by
\begin{equation}
\varepsilon \lbrack \phi (x)]=%
\begin{cases}
\lambda _{1}^{2}a_{1}^{4}\mathrm{{sech}^{2}(\lambda _{1}a_{1}(x-x_{1}))} &
\quad ;-\infty <x<x_{1} \\
&  \\
\frac{2\lambda _{1}^{2}a_{1}^{4}}{B_{1}}(b_{1}^{2}+B_{1})\cn^{2}\bigg(%
h_{1}(x)\bigg|\frac{b_{1}^{2}+B_{1}}{2B_{1}}\bigg)\dn^{2}\bigg(h_{1}(x)\bigg|%
\frac{b_{1}^{2}+B_{1}}{2B_{1}}\bigg) & \quad ;x_{1}\leq x<x_{2} \\
&  \\
\frac{2\lambda _{1}^{2}a_{1}^{4}}{B_{2}}(b_{2}^{2}+B_{2})\cn^{2}\bigg(%
h_{2}(x)\bigg|\frac{b_{2}^{2}+B_{2}}{2B_{2}}\bigg)\dn^{2}\bigg(h_{2}(x)\bigg|%
\frac{b_{2}^{2}+B_{2}}{2B_{2}}\bigg) & \quad ;x_{2}\leq x<x_{3} \\
&  \\
\lambda _{1}^{2}a_{1}^{4}sech^{2}\bigg(\lambda _{1}\frac{a_{1}^{2}}{a_{2}}%
(x-x_{3})\bigg) & \quad ;x_{3}\leq x<+\infty ,%
\end{cases}
\label{energy_phi_assimetrico}
\end{equation}%
where we defined $h_{1}(x)=\delta _{1}+\sqrt{2/B_{1}}\lambda
_{1}a_{1}^{2}(x-x_{1})$ and $h_{2}(x)=\delta _{2}+\sqrt{2/B_{2}}\lambda
_{1}a_{1}^{2}(x-x_{2})$.

\section{Solution of the second smooth model with symmetric local vacua:}

\indent Now, we will treat the case of the potential (\ref{potecial_2}). We
are again interested in solutions connecting a negative vacuum with a
positive one, so that they shall obey the following boundary conditions
\begin{equation}
\lim_{x\rightarrow \pm \infty }\phi (x)=\pm (a+n\pi /\mu ),
\label{boundary_2}
\end{equation}%
and using the very same argument of the first model, one is lead to conclude
that $\phi (x)$ shall be a monotonically increasing function. Despite the
fact that $V(\phi )$ is written in terms of only three regions, it will be
convenient to divide the axis $\phi $ in $2n+2$ regions, each one localized
between a local maximum and the next adjacent maxima. In each potential
region we associate a set $\Phi _{i}$, and this leads us to%
\begin{align}
\Phi _{1}& =\{\phi \quad |\phi <-n\pi /\mu \},  \notag \\
\Phi _{2}& =\{\phi \quad |-n\pi /\mu \leq \phi <(1-n)\pi /\mu \},  \notag \\
& \vdots  \notag \\
\Phi _{i}& =\{\phi \quad |(i-n-2)\pi /\mu \leq \phi <(i-n-1)\pi /\mu \},
\notag \\
& \vdots  \notag \\
\Phi _{2n+2}& =\{\phi \quad |n\pi /\mu \leq \phi \}.
\end{align}%
As it was done in the case of the first model, we associate a set $X_{i}$ to
each $\Phi _{i}$, and we get that $\phi :X_{i}\longrightarrow \Phi _{i}$.
This allow us to define the sets $X$ as
\begin{align}
X_{1}& =\{x\quad |x_{1}>x\},  \notag \\
X_{2}& =\{x\quad |x_{1}\leq x<x_{2}\},  \notag \\
& \vdots  \notag \\
X_{i}& =\{x\quad |x_{i-1}\leq x<x_{i}\},  \notag \\
& \vdots  \notag \\
X_{2n+2}& =\{x\quad |x_{2n+1}\leq x\},
\end{align}%
where $\phi (x_{i})=(i-n-1)\pi /\mu $.

\indent\textbf{Solution in the region 1:} In this region the solution $%
\phi:X_{1}\longrightarrow\Phi_{1}$, comes from the equation
\begin{equation}
\frac{1}{2}\phi^{\prime2}=\frac{\lambda^{2}}{2}(a^{2}-(\phi+n\pi/%
\mu)^{2})^{2}.
\end{equation}
We can identify the above equation with the one appearing in (\ref{reg_1})
simply by doing $b=\pi/\mu$ and $\phi(x)$ can be obtained from (\ref%
{solution_1}), giving
\begin{equation}
\phi(x)=a\tanh(\lambda a(x-x_{1}))-n\pi/\mu\quad;-\infty<x<x_{1}.
\label{solution_1_2}
\end{equation}
It is easy to verify that it obeys the correct boundary conditions.

\indent\textbf{Solution in the 2n+2 region:} In this last region $%
\phi:X_{2n+2}\longrightarrow\Phi_{2n+2}$ the differential equation is such
that
\begin{equation}
\frac{1}{2}\phi^{\prime2}=\frac{\lambda^{2}}{2}(a^{2}-(\phi-n\pi/%
\mu)^{2})^{2}.
\end{equation}
Now, choosing $b=\pi/\mu$, the solution of (\ref{solution_2}) leads us to
\begin{equation}
\phi(x)=a\tanh(\lambda a(x-x_{2}))+n\pi/\mu\quad;x_{2}\leq x<+\infty.
\end{equation}

\indent\textbf{Intermediary solutions: }In these regions the differential
equations look like%
\begin{equation}
\frac{1}{2}\phi ^{\prime 2}=\frac{\alpha ^{2}}{2}\left[ B^{2}+\frac{%
1+(-1)^{n}\cos \left( \mu \phi \right) }{2}\right] .  \label{mod_2}
\end{equation}%
Again, we are looking for a solution of the type $\phi
(x):X_{i}\longrightarrow \Phi _{i}$. However, the process of solution will
be the same in all the intermediary regions, and it will be necessary only
to adjust the appropriate boundary conditions in each of them. First of all,
we will work with an even $n$. In this case, we can rewrite the differential
equation (\ref{mod_2}) as
\begin{equation}
\frac{d\phi }{dx}=\alpha \sqrt{B^{2}+1-\sin ^{2}(\mu \phi /2)}.
\label{mod_2_2}
\end{equation}%
Performing the transformation of variable $\varphi =\mu \phi /2$, one can
integrate the above equation in the following manner
\begin{equation}
\int_{\varphi (x_{i-1})}^{\varphi (x)}\frac{d\varphi ^{\prime }}{\sqrt{%
1-m\sin ^{2}\varphi ^{\prime }}}=\frac{\mu \lambda a^{2}}{2}%
\int_{x_{i-1}}^{x}dx^{\prime }=\frac{\mu \lambda a^{2}}{2}(x-x_{i-1}),
\label{int_2}
\end{equation}%
where we defined $m\equiv 1/(1+B^{2})$ (with $m\in \lbrack 0,1]$) and used
the equation (\ref{constantes_2}) in order to eliminate the parameter $%
\alpha $ in terms of $\lambda a^{2}$. The left-hand side of this last
equation represents an elliptic integral, and (\ref{elliptic}) allows us to
write
\begin{equation}
\sin (\mu \phi (x)/2)=\sn\bigg(\delta _{i}+\frac{\mu \lambda a^{2}}{2}%
(x-x_{i-1})\bigg|\frac{1}{1+B^{2}}\bigg),
\end{equation}%
where $\delta _{i}=\sn^{-1}[\sin (\mu \phi (x_{i-1})/2)]$. On the other
hand, by using $\phi (x_{i})=(i-n-1)\pi /\mu $, it can be shown that
\begin{equation}
\sin (\mu \phi (x_{i-1})/2)=(-1)^{\frac{n}{2}+1}\sin (i\pi /2)  \label{seno}
\end{equation}

At this point we shall be careful, since the field $\phi (x)$ depends on the
function $\arcsin (x)$ which, by its turn, must have its image and dominion
very well specified, due its multivalence. In order to avoid such kind of
problem, we specify
\begin{equation}
\arcsin :[-1,1]\longrightarrow \lbrack -\pi /2,+\pi /2].
\end{equation}%
On the other hand, the boundary conditions are assured since the equation (%
\ref{mod_2_2}) is invariant under transformations of the type $\phi
(x)\rightarrow \phi (x)+k\pi $ (valid for integer values of $k$). Thus, we
can write
\begin{equation}
\phi (x)=\frac{2}{\mu }\arcsin \bigg[\sn\bigg(\delta _{i}+\frac{\mu \lambda
a^{2}}{2}(x-x_{i-1})\bigg|\frac{1}{1+B^{2}}\bigg)\bigg]-\frac{2k\pi }{\mu }%
\quad ;x_{i-1}\leq x<x_{i}.
\end{equation}%
Now, using the boundary condition $\phi (x_{i})=(i-n-1)\pi /\mu $ in the
above equation, we get
\begin{equation}
\phi (x_{i-1})=\frac{2}{\mu }\arcsin \bigg[\sn\bigg(\delta _{i}\bigg|\frac{1%
}{1+B^{2}}\bigg)\bigg]-\frac{2k\pi }{\mu }=\frac{i-n-2}{\mu }\pi .
\end{equation}%
Using the definition of $\delta _{i}$ and the equation (\ref{seno}), we can
conclude that~
\begin{equation}
k=%
\begin{cases}
\frac{n+2-i}{2} & \text{; for even $i$ } \\
\frac{n+2+[(-1)^{\frac{n}{2}+1}-1]i}{2} & \text{; for odd $i$ }.%
\end{cases}
\label{k}
\end{equation}%
Before finishing the discussion of this section, we shall determine the
relative position of the $x_{i}$. From the boundary condition in $\phi
(x_{i})$, we can write
\begin{equation}
\phi (x_{i})=\frac{2}{\mu }\arcsin \bigg[\sn\bigg(\delta _{i}+\frac{\mu
\lambda a^{2}}{2}(x_{i}-x_{i-1})\bigg|\frac{1}{1+B^{2}}\bigg)\bigg]-\frac{%
2k\pi }{\mu }=\frac{i-n-1}{\mu }\pi .
\end{equation}%
From the definition of $\delta _{i}$ and using the equation (\ref{k}), we
arrive at
\begin{align}
& \sn\bigg(\frac{\mu \lambda a^{2}}{2}(x_{i}-x_{i-1})\bigg|\frac{1}{1+B^{2}}%
\bigg)=1\quad \text{; for even $i$ },  \notag  \label{xs} \\
& \delta _{i}+\frac{\mu \lambda a^{2}}{2}(x_{i}-x_{i-1})=0\quad \text{; for
odd $i$ }.
\end{align}%
Through the above equation it can be shown that
\begin{align}
& x_{3}-x_{1}=x_{5}-x_{3}=x_{7}-x_{7}=...=x_{i}-x_{i-2};\quad \text{; for
even $i$ },  \notag \\
& x_{4}-x_{2}=x_{6}-x_{4}=x_{8}-x_{6}=...=x_{i}-x_{i-2};\quad \text{; for
odd $i$ }.
\end{align}%
Again, once the value of $x_{1}$ is defined, we can determine $x_{2}$ and $%
x_{3}$ through (\ref{xs}) and the remaining junction points are given by the
above equations.

\indent In general, the solution for the multi-kink field configuration of
this model can be written in the form
\begin{equation} \label{phi_2}
\phi (x)=%
\begin{cases}
a\tanh (\lambda a(x-x_{1}))-n\pi /\mu & \quad ;-\infty <x<x_{1} \\
&  \\
\frac{2}{\mu }\arcsin \bigg[\sn\bigg(\delta _{i}+\frac{\mu \lambda a^{2}}{2}%
(x-x_{i-1})\bigg|\frac{1}{1+B^{2}}\bigg)\bigg]-\frac{2k\pi }{\mu } & \quad
;x_{i-1}\leq x<x_{i} \\
&  \\
a\tanh (\lambda a(x-x_{2n+2}))+n\pi /\mu & \quad ;x_{2n+2}\leq x<+\infty .%
\end{cases}%
\end{equation}%
As expected, this solution has the same profile as the one obtained for the
first model which is appearing in the Figure 2. All the above discussion was
done for the case with even $n$. However, the same procedure can be used for
the case with odd $n$ and the result will have the same appearance, and due
to this we will not present the details here. The energy density of this
soliton can also be calculated by using the equation (\ref{energy}), and it
presents a very similar profile of the one corresponding to the first model,
which can be seen in the Figure 2.

\section{Topological Properties:}

In this section we will consider some topological properties for the models proposed in this paper. It is well know that the usual kink-like configurations are topological solutions and, as a consequence, we may use some index in order to classify such solutions with respect to the topological features. The so called topological charge is an example of topological index often used in the literature. Here, we show that the topological charge it is well defined for the models presented here, and we will calculate this charge for both models.\\
\indent We have defined a topological current in usual way \cite{Rajaraman}, namely
\begin{equation}
j^{\mu} = \varepsilon^{\mu \nu}\partial_{\nu} \phi ,
\end{equation}
where $\varepsilon^{\mu \nu}$ is the Levi-Civita symbol. It is not difficult to conclude that a conservation law follows from the definition of $j^{\mu}$, in fact we have $\partial_{\mu}j^{\mu} = 0$ as a consequence of the antisymmetry of $\varepsilon^{\mu \nu}$. Note that both the topological charge and its conservation law are well defined, since we have ensured that first and the second derivatives of the scalar fields solutions are continuous at any point. The topological charge may be defined in terms of $j^{\mu}$ in the following way
\begin{equation}
Q = \int_{-\infty}^{\infty} dx \, j^0.
\end{equation}
Note that $j^0 = \varepsilon^{0 \nu}\partial_{\nu} \phi = \partial \phi/ \partial x$, thus, after integration we obtain the following result
\begin{equation}
Q = \phi(x= +\infty) - \phi(x= - \infty).
\end{equation}
\indent For the first model considered in this work, whose solution is given by eq. (\ref{phi_1}), we have the asymptotic values for the scalar field given by
\begin{equation}
\phi(x= +\infty) = a + n b \quad \textmd{and} \quad \phi(x= -\infty) = -a -  n b.
\end{equation}
Therefore, the topological charge for the first model is given by
\begin{equation}
Q = 2a + 2 nb.
\end{equation}
It is interesting to note that the topological charge obtained for the usual $\phi^4$ theory (which may be obtained with $n=0$ in eq. (\ref{10})) is given by $Q_0 = 2a$, thus, we may rewrite the case with arbitrary $n$ as follows
\begin{equation}
Q_n = Q_0 + 2 n b.
\end{equation}

\indent Repeating the same procedure for the second model, whose solution is given by eq. (\ref{phi_2}), we obtain $\phi(x= +\infty) = a + n \pi/\mu$ and $\phi(x= -\infty) = -a - n \pi/\mu$. Therefore, in this case the topological charge is given by
\begin{equation}
Q_n = 2a + 2 n \pi/\mu = Q_0 + 2 n \pi/\mu .
\end{equation}
\indent We may note that in both cases the topological charge increases linearly with the number of local vacua as well as with the separation of them .
\bigskip

\section{Conclusions:}

\bigskip In this work, we have introduced a method which can be
systematically used to obtain analytical multi-kink configurations which
come from very smooth stepwise scalar field potentials. The approach was
presented through three examples, and their corresponding typical
triple-kink and energy densities profiles can be seen in the Figures 2 and
3. In fact, as it can be observed from the case studied in the Introduction
section, these multi-kinks profiles can be constructed in the case of
potentials similar to the DQ model and its generalizations (see Figure 1).
However, those potentials present discontinuity in their derivative at the
junction points, which do not happens in the cases we have introduced in
this work. Among the possible applications of our results, we are presently
interested in the possibility of constructing multi-brane-world scenarios.

\bigskip

\textbf{Acknowledgements: }The authors thanks to CNPq and FAPESP for partial
financial support.

\newpage
\begin{figure*}[tbp]
\centering
\includegraphics{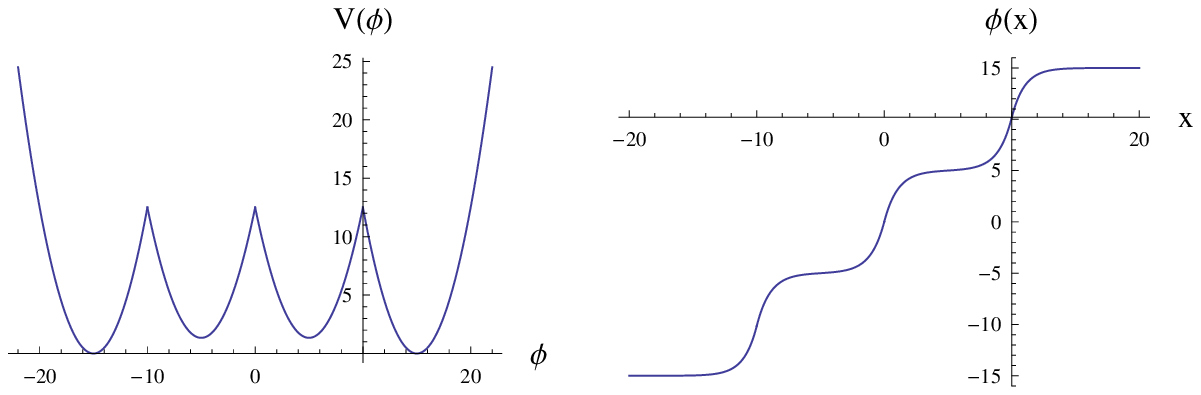}
\caption{Stepwise Quadratic Potential presenting two global and two local
vacua and the corresponding triple-kink ($a=10, \, b_1=b_4=0,\, \protect%
\lambda_1=1$ and $b_2=b_3=3$)}
\label{fig1}
\end{figure*}

\begin{figure*}[tbp]
\centering
\includegraphics{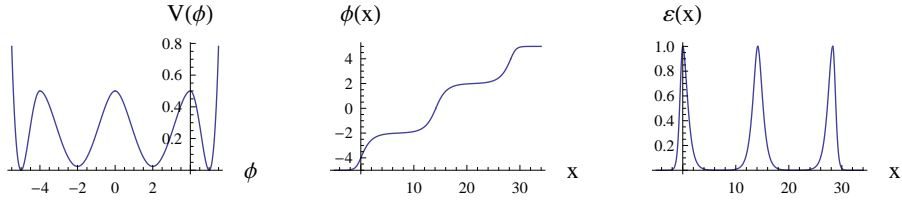}
\caption{Smooth Symmetric Potential presenting two global and two local
vacua, its triple-kink and energy density ($a=\protect\lambda=1,\, n=2,\,
b=2 $ and $B=4.1$)}
\label{fig2}
\end{figure*}

\begin{figure*}[tbp]
\centering
\includegraphics{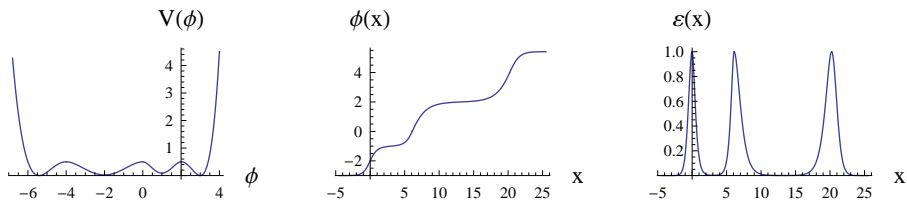} \label{fig3}
\caption{Smooth Asymmetric Potential presenting two global and two local
vacua, its triple-kink and energy density($a_1=\protect\lambda_1=b_1=1,\,
b_2=2,\, \protect\lambda_2=1/2, \, a_2=\protect\sqrt{2}, \, B_1=1.1$ and $%
B_2=4.2$)}
\end{figure*}

\end{document}